\documentstyle[preprint,pra,aps,eqsecnum,dina4]{revtex}

\begin{document}
\title{Cat States and Single Runs for the Damped Harmonic
 Oscillator}
\author{Peter Goetsch, Robert Graham, and Fritz Haake}
\address{Fachbereich Physik, Universit\"at-Gesamthochschule Essen,
 45117 Essen, Germany}
\maketitle

\begin{abstract}
We discuss the fate of initial states of the cat type for the damped
harmonic oscillator, mostly employing a linear version of the
stochastic Schr\"odinger equation. We also comment on how such cat
states might be prepared and on the relation of single realizations
of the noise to single runs of experiments.
\end{abstract}

\section{Introduction}
\label{sec:intro}
The conventional quantum description of open systems employs density
matrices and their nonunitary evolution equations, i.~e. master
equations. An equivalent description can be given in terms of wave
functions, at the expense of introducing  damping and noise terms in
the Schr\"odinger equation (see e.~g.~\cite{Pe,Gi,Di,DaCo,Gi2,GiPer%
,DaCaMo,DuZoRi,GaPaZo,DuPaZoGa,MoCaDa,WiMi,Ca,GoGr1,GhPeRi,Be,BeSt%
,BaBe,Ho,Ba,GoGr2,GoGr3}). For a given open system, both the master
equation and the corresponding stochastic Schr\"odinger equation can
be derived by treating the interaction, usually weak, with an
environment.

Whether one deals with a master equation or the equivalent stochastic
Schr\"odinger equation is mostly a matter of convenience. Less work
is involved in determining $N$ components of a state vector for a
single realisation of the noise process rather than $N^2$ elements of
a density matrix; however, many realisations of the noise must be
followed in order to establish the full information stored in the
density matrix.

Apart from occasional computational convenience the stochastic
Schr\"odinger equation has a certain conceptional appeal. Recent
experiments on individual ions or atoms stored in electromagnetic
traps have revived the old question about what quantum mechanics has
to say about the course of a single run of an experiment. The
so-called quantum jumps of an atom into or out of a metastable state
and the accompanying random-telegraph modulation of a fluorescence
signal, for instance, strengthen our interest in a formulation of
dissipative quantum dynamics where random jumps are naturally
reflected in the time dependence of the wave function.

The stochastic Schr\"odinger equation is particularly welcome to some
researchers interested in quatum measurement theory. The seemingly
scandalous effective collapse of a superposition to a mixture upon
measurement is by now rather well understood, even on the basis of
exactly solvable models \cite{HaZu,HaWa,Zu}. However, it is still
esthetically satisfying to see the wavefunction of a pointer variable
evolve, in each realization of the noise, to one particular
distinguished displacement corresponding to one eigenvalue of the
measured observable of some microsystem; and to see many independent
realisations of the noise build up the relative frequencies of the
possible outcomes.

Other peculiarities of quantum mechanics, like Schr\"odinger's cat
and the EPR paradox, appear under an interesting new perspective in
the description by a stochastic Schr\"odinger equation. The fate of
Schr\"odinger's cat was recently investigated numerically in a simple
model in this manner \cite{GarKni}. Previous treatments of
such problems using the density matrix \cite{WaMi,Ga} were thus
complemented in rather more intuitive ways.

We here take up the cat problem again. We present a rigorous solution
of a stochastic Schr\"odinger equation for the damped harmonic
oscillator, starting from an initial state of the cat type. It is
most gratifying to see how the wavevector chooses randomly between
the two possible and ``macroscopically'' distinguishable locations.
The choice is made on the so-called {\it decoherence} time scale
$t_{\text{dec}}$ which is shorter than the mechanical damping time by
a factor measuring the difference between the two initial locations.
For each realization of the noise, jumps between the two locations
become exceedingly unlikely after several units of $t_{\text{dec}}$.

Noise can be accounted for in the stochastic Schr\"odinger equation
in different but essentially equivalent ways. One possibility is to
let the noise represent vacuum and/or thermal fluctuations of the
environment; it is then natural to speak of input noise.
Alternatively, we may let the noisy part of the output of the system
appear as stochastic driving. Both forms of the noise are of course
related to one another, as described by the input-output formalism of
Collet and Gardiner \cite{CoGa}. Depending on which form of the noise
is chosen in the stochastic Schr\"odinger equation, the latter takes
a linear or nonlinear appearance. The equivalence of the two
formulations, recently shown by Ghirardi, Pearle and Rimini
\cite{GhPeRi}, by Belavkin and Staszewski \cite{BeSt} and by Goetsch
and Graham \cite{GoGr2,GoGr3}, will be illustrated again below.

\section{Linear evolution from a cat state}
\label{sec:linear}
We first discuss the solution of the linear stochastic Schr\"odinger
equation \cite{GhPeRi,Be,BeSt,BaBe,Ho,Ba,GoGr2,GoGr3}
\begin{equation}
\label{eq:2-1}
 d|\psi_\theta (t)\rangle =
  \left\{
   -\frac{\gamma}{2}a^\dagger a dt +\sqrt{\gamma}
     ae^{-i\varphi}d\theta(t)
  \right\}
  |\psi_\theta (t)\rangle\,.
\end{equation}
Here $d\theta(t)$ is a real noise increment independent of its
predecessors in time; otherwise, $d\theta(t)$ need not be specified
at this point. The index $\theta$ on the wave vector is meant to
indicate the contingency of that vector on the noise history:
$|\psi_\theta (t)\rangle$ is a functional of the noise increments
arisen in the past but independent of the one arising at present,
$d\theta (t)$. To include the description of homodyning with a local
oscillator of constant phase $\varphi$ we have allowed for the phase
factor $e^{-i\varphi}$ \cite{WiMi,GoGr3}. Obviously, the generator
appearing on the rhs of the stochastic Schr\"odinger equation is
non-Hermitian for which reason the normalization of the vector
$|\psi_\theta (t)\rangle$ is not preserved in time.

Due to its linearity in the state vector $|\psi_\theta (t)\rangle$
and the bilinearity of the generator in $a$ and $a^\dagger$, the
stochastic Schr\"odinger equation (\ref{eq:2-1}) can be solved
rigorously. For instance, initial coherent states $|\pm\alpha
\rangle$, defined as eigenstates of $a$ with eigenvalues
$\pm\alpha$, evolve into \cite{GoGr2,GoGr3}
\begin{equation}
\label{eq:2-2}
c_\pm(t)|\pm\alpha e^{-\frac{\gamma}{2}t}\rangle,
\end{equation}
i.~e.~coherent states of amplitudes
$\pm\alpha e^{-\frac{\gamma}{2}t}$ multiplied with stochastic
factors of non-unit modulus,
\begin{equation}
\label{eq:2-3}
c_\pm(t)=\exp\Bigg\{
                \frac{1}{2}\left(\alpha^2+|\alpha|^2\right)
                           \left(e^{-\gamma t}-1\right) \pm
                \alpha\sqrt{\gamma}\int_0^t e^{-\frac{\gamma}{2}s}
                 d\theta(s)
             \Bigg\}.
\end{equation}
Of special interest to us will be an initial superposition of two
coherent states,
\begin{equation}
\label{eq:2-4}
 |\psi(t=0)\rangle =
 \frac{1}{\sqrt{2}}\left(|\alpha\rangle+|-\alpha\rangle\right).
\end{equation}
For large values of the amplitudes, $|\alpha|\gg 1$, for which the
constituents $|\pm\alpha\rangle$ become ``macroscopically
distinguishable'', such a superposition is customarily called a
{\it Schr\"odinger cat state}. Invoking the superposition principle
we obtain the state originating from the superposition (\ref{eq:2-4})
as
\begin{equation}
\label{eq:2-5}
|\psi_\theta (t)\rangle =
      \frac{1}{\sqrt{2}} \left(
         c_+(t)|\alpha e^{-\frac{\gamma}{2}t}\rangle
       + c_-(t)|-\alpha e^{-\frac{\gamma}{2}t}\rangle \right).
\end{equation}
We can now address the mean of the quadrature component
\begin{equation}
\label{eq:2-6}
X_\varphi=\frac{1}{2}\left(ae^{-i\varphi}+a^\dagger e^{i\varphi}
\right)
\end{equation}
with respect to the normalized version of the time-evolved cat state
(\ref{eq:2-5}). To avoid unnecessary complication we shall henceforth
take the amplitude $\alpha$ as real and thus get
\begin{eqnarray}
\label{eq:2-7}
\langle X_\varphi(t)\rangle_\theta
&=& {\cal N_\theta}^{-1}\langle\psi_\theta(t)|X_\varphi|\psi_\theta
    (t)\rangle\nonumber\\
&=& {\cal N_\theta}^{-1}(\alpha/2)e^{-\frac{\gamma}{2}t}
     \Bigg\{\left(|c_+|^2-|c_-|^2\right)\cos\varphi \nonumber\\
& & \phantom{-}+ i(c_+^\star c_- -c_+c_-^\star)
      \langle\alpha e^{-\frac{\gamma}{2}t}
           |-\alpha e^{-\frac{\gamma}{2}t}\rangle
\sin\varphi\Bigg\},\nonumber\\
{\cal N_\theta}
&=& \langle\psi_\theta(t)|\psi_\theta(t)\rangle.
\end{eqnarray}
Clearly, this mean displacement vanishes initially, due to the
symmetry of the initial state, and for large times, $t\gg 1/\gamma$,
due to the decay of $|\psi_\theta(t)\rangle$. However, at
intermediate times we encounter a nonvanishing functional of the
noise $\theta(t)$. The dependence of $\langle
X_\varphi(t)\rangle_\theta$
on the phase of the local oscillator is interesting:  For
$\varphi=0$ the overlap of the two normalized coherent states
$|\pm\alpha e^{-\frac{\gamma}{2}t}\rangle$,
\begin{equation}
\label{eq:2-8}
\langle\alpha e^{-\frac{\gamma}{2}t}
     |-\alpha e^{-\frac{\gamma}{2}t}\rangle =
\exp\left(-2\alpha^2e^{-\gamma t}\right)\,,
\end{equation}
enters only through the normalization factor ${\cal N}_\theta$; for
$\varphi =\pi/2$, on the other hand, the presence of that overlap is
essential for the transient appearance of a finite mean
$\langle X_{\pi/2}(t)\rangle_\theta$, the exponential smallness of
$\langle\alpha|-\alpha\rangle=\exp(-2\alpha^2)$ notwithstanding.
Having pointed out this delicate dependence on the phase $\varphi$
we proceed to setting $\varphi=0$ and writing $X$ instead of $X_0$,
for simplicity.

We shall mostly be concerned with times small compared to the
``mechanical'' relaxation time $1/\gamma$. In that regime the mean
displacement takes the simple form
\begin{equation}
\label{eq:2-9}
\langle X(t)\rangle_\theta =
  \alpha\tanh\left(2(\alpha\sqrt{\gamma})\theta(t)\right),
    \quad t\ll 1/\gamma;
\end{equation}
this results by approximating as
$e^{-\frac{\gamma}{2}t}\approx 1$ and dropping the exponentially
small overlap $\langle\alpha|-\alpha\rangle$. We should note that the
conditional mean $\langle X(t)\rangle_\theta$, being a functional of
the noise $\theta(s)$ in $0\le s< t$, is a random quantity itself. It
fluctuates from one realization of the process $\theta(t)$ to another
one, each of which one might feel tempted to associate with a
particular run of an experiment.

In order to compute full quantum averages of oscillator observables
$Y$ like the displacement $X$ we still have to average over the noise
$\theta(t)$ which we now need to specify. We want to describe an
oscillator weakly interacting with a zero-temperature reservoir. By
invoking the usual Born and Markov approximation, the stochastic
Schr\"odinger equation was derived in \cite{BeSt,GoGr2,GoGr3},
together with the prescription for calculating quantum means
\begin{equation}
\label{eq:2-10}
\langle Y(t)\rangle=\int d\mu_t^W(\{\theta\})\,
  \langle\psi_\theta(t)|Y|\psi_\theta(t)\rangle.
\end{equation}
We here encounter the Wiener measure $d\mu_t^W(\{\theta\})$ for
finding a realization $\theta(s)$ of the noise during the time
interval $0\le s< t$. The result (\ref{eq:2-10}) allows us to think
of $\theta(t)$ as the vacuum fluctuations of the reservoir forced on
the oscillator as an ``input''.

An alternative useful way of thinking of the noise is suggested
by introducing the normalized version of the state
$|\psi_\theta(t)\rangle$
\begin{equation}
\label{eq:2-11}
\langle Y(t)\rangle_\theta =
 \langle\psi_\theta(t)|Y|\psi_\theta(t)\rangle/
  \langle\psi_\theta(t)|\psi_\theta(t)\rangle
\end{equation}
and rewriting (\ref{eq:2-10}) as
\begin{eqnarray}
\label{eq:2-12}
\langle Y(t)\rangle &=& \int d\mu_t^W(\{\theta\})\,
  \langle\psi_\theta(t)|\psi_\theta(t)\rangle\langle Y(t)
  \rangle_\theta\nonumber\\
 &\equiv & \int d\mu_t(\{\theta\})\,\langle Y(t)\rangle_\theta.
\end{eqnarray}
We may interpret
\begin{equation}
\label{eq:2-13}
 d\mu_t(\{\theta\})=d\mu_t^W(\{\theta\})
  \langle\psi_\theta(t)|\psi_\theta(t)\rangle
\end{equation}
as a noise measure as well \cite{GhPeRi,Be,BeSt,BaBe,Ho,Ba,GoGr2,%
GoGr3}; indeed, by choosing $Y=1$ in (\ref{eq:2-12}) we have the
correct normalization $\int d\mu_t(\{\theta\})=1$. As was noted in
\cite{BeSt,GoGr2,GoGr3} one may regard the random process with the
measure $d\mu_t(\theta)$ as the output of the oscillator driven by
vacuum fluctuations, i.~e.~a Wiener process $d\xi(t)$ as input,
\begin{equation}
\label{eq:2-14}
 d\theta(t)=2\sqrt{\gamma}\langle X_\varphi(t)\rangle_\theta dt
 +d\xi(t).
\end{equation}
Here, we have intentionally restored the index $\varphi$ on the
quadrature component $X_\varphi$ since the input-ouput connection
(\ref{eq:2-14}) remains in fact valid for arbitrary values of the
phase of the local oscillator.

The two interpretations of the noise suggest different possibilities
of calculating the means $\langle Y(t)\rangle$ by numerical
simulation. One, and in fact the simpler one, is to
generate random numbers with the Wiener measure $d\mu_t^W$, use them
in evaluating $|\psi_\theta(t)\rangle$, and then employ
(\ref{eq:2-10}). According to the second strategy one would again
construct realizations of the Wiener process and a set of ensuing
$|\psi_\theta(t)\rangle$ but then proceed to the new measure
$d\mu_t$ and finally average as required in (\ref{eq:2-12}). We
should stress that the expectation value $\langle X(t)\rangle_\theta$
in the definition (\ref{eq:2-14}) is meant in the sense of the
second strategy, i.~e.~conditioned on the measure $d\mu_t$ for the
past history $\theta(s)$, $s\le t$.

We now propose a closer inspection of the measure $d\mu_t(\theta)$
for times $0\le t\ll 1/\gamma$. As already noted above we may 

then drop the overlap $\langle\alpha e^{-\frac{\gamma}{2}t}|
-\alpha e^{-\frac{\gamma}{2}t}\rangle\approx\langle\alpha|-\alpha
\rangle$, thus obtaining
\begin{eqnarray}
\label{eq:2-15}
\langle\psi_\theta(t)|\psi_\theta(t)\rangle &=&
  \frac{1}{2}\left(|c_+|^2+|c_-|^2\right)\nonumber\\
    &=& e^{-2\alpha^2\gamma t}\cosh
     \left[2\alpha\sqrt{\gamma}\int_0^t d\theta(s)\right],\quad
      t\ll 1/\gamma.
\end{eqnarray}
In writing out the measure $d\mu_t(\{\theta\})$ more explicitly it
is convenient to discretize the time as $t_n=n\Delta t$,
$n=0\dots N$, $t_N=t$ and to consider increments $\Delta_n\theta$
such that $\theta(t_n)=\sum_{i=0}^{n-1}\Delta_i\theta$. Assuming
$\theta(0)=0$ we then have
\begin{eqnarray}
\label{eq:2-16}
 d\mu_t(\{\theta\}) &=& \frac{1}{2}
  \Bigg[\prod_{i=0}^{N-1}\frac{d\Delta_i\theta}{\sqrt{2\pi\Delta t}}
    \,\exp\left(
    -\left(\Delta_i\theta-2\alpha\sqrt{\gamma}\Delta t\right)^2
    /2\Delta t\right)\nonumber\\
   &&\quad
     +\prod_{i=0}^{N-1}\frac{d\Delta_i\theta}{\sqrt{2\pi\Delta t}}
      \,\exp\left(
      -\left(\Delta_i\theta+2\alpha\sqrt{\gamma}\Delta t\right)^2
    /2\Delta t\right)\Bigg].
\end{eqnarray}
When averaging the mean displacement (\ref{eq:2-9}) with this
measure we may first reduce $d\mu_t(\{\theta\})$ to the marginal
density of the single variable $\theta(t)\equiv\theta$
\begin{eqnarray}
\label{eq:2-17}
P(\theta,t) &=& \int d\mu_t(\{\theta\})\,\delta\left(\theta-
  \sum_{i=0}^{N-1}\Delta_i\theta\right)\nonumber\\
 &=& \frac{1}{2}\{P^{(+)}(\theta,t)+P^{(-)}(\theta,t)\},\nonumber\\
P^{(\pm)} &=& \frac{1}{\sqrt{2\pi t}}
  \exp\{-(\theta\mp 2(\alpha\sqrt{\gamma}) t)^2/2t\}.
\end{eqnarray}
The two Gaussians here appearing are not resolved from one another
as long as their separation $4(\alpha\sqrt{\gamma})t$ is smaller than
their width $\sqrt{t}$. However, as soon as
\begin{equation}
\label{eq:2-18}
 t\gg \frac{1}{2\gamma\alpha^2}\equiv t_{\text{dec}}
\end{equation}
the density $P(\theta,t)$ has two non-overlapping peaks. Since the
so-called decoherence time $1/2\gamma\alpha^2$ \cite{WaMi,Ga} is,
for strongly excited coherent states $(\alpha^2\gg 1)$, much shorter
than the mechanical life time of excitations $1/\gamma$, the
separation in question takes place well within the range of validity
$(t\ll 1/\gamma)$ of the various simplifications made in arriving at
(\ref{eq:2-17}). We may conclude that the displacement of the
oscillator has an overwhelming likelihood, at all times in the
interval $1/2\gamma\alpha^2\ll t\ll 1/\gamma$, to take on either one
or the other of two ``macroscopically'' distinguishable values rather
than their vanishing ensemble average. Fig.~\ref{fig1} shows the
density $P(\theta,t)$ for various moments of time.

Moreover, one intuitively expects jumps between the two
macroscopically distinct values of $\langle X(t)\rangle_\theta$ to
become less and less frequent as $t$ begins to exceed the decoherence
time $t_{\text{dec}}$. In order to substantiate this expectation we
consider the joint density
\begin{equation}
\label{eq:2-19}
P(\theta,t;\theta',t')=\int d\mu_{t'}(\{\theta\})\,
  \delta\left(\theta-\sum_{i=1}^{N-1}\Delta_i\theta\right)
   \delta\left(\theta'-\sum_{j=1}^{N'-1}\Delta_j\theta\right)
\end{equation}
with $t'=N'\Delta t\ge t=N\Delta t$. By noting
$\theta'-\theta=\sum_{i=N}^{N'-1}\Delta_i\theta$ and doing the
Gaussian integrals we arrive at
\begin{eqnarray}
\label{eq:2-20}
 P(\theta,t;\theta',t') &=& \frac{1}{2}
  \bigg\{P^{(+)}(\theta,t)P^{(+)}(\theta'-\theta,t'-t)\nonumber\\
 &&\qquad +P^{(-)}(\theta,t)P^{(-)}(\theta'-\theta,t'-t)\bigg\}.
\end{eqnarray}

It is interesting to realize how the double-peak structure
(\ref{eq:2-17}) of the single-point density generalizes to the
two-point density (\ref{eq:2-20}). In particular, the absence of
cross terms $P^{(+)}P^{(-)}$ suggests that there is no
interference between what happens near
$\langle X\rangle_\theta\approx+\alpha$ and near
$\langle X\rangle_\theta\approx-\alpha$. For a quantitative estimate
of the frequency of jumps between the two branches we measure
$t$ and $t'$ in units of the decoherence time
$(t=\tau/2\gamma\alpha^2)$ and evaluate the probability of finding
$\theta$ within a width $\sqrt{t}$ around the peak at
$+2(\alpha\sqrt{\gamma})t$, and $\theta'$ within a width $\sqrt{t'}$
around the peak at $-2(\alpha\sqrt{\gamma})t'$,
\begin{eqnarray}
\label{eq:2-21}
  p(\tau,\tau')&=&\text{prob}\Bigg\{
   \theta\left(\frac{\tau}{2\gamma\alpha^2}\right)\in
    \left[\frac{1}{\alpha\sqrt{\gamma}}\left(\tau-\sqrt{\tau/8}
    \right),\frac{1}{\alpha\sqrt{\gamma}}\left(\tau+\sqrt{\tau/8}
    \right)\right]\nonumber\\
  &&\quad{\text{and }}\ \theta'\left(\frac{\tau'}{2\gamma\alpha^2}
    \right)\in\left[\frac{1}{\alpha\sqrt{\gamma}}
    \left(\tau'-\sqrt{\tau'/8}\right),\frac{1}{\alpha\sqrt{\gamma}}
    \left(\tau'+\sqrt{\tau'/8}\right)\right]\Bigg\}.
\end{eqnarray}
We immediately obtain this ``jump probability'' as
\begin{equation}
\label{eq:2-22}
 p(\tau,\tau')=\frac{1}{2\pi}
 \left\{\int_{\text{I}_1}dx\,e^{-x^2}\int_{\text{I}_2}dy\,e^{-y^2}
  +\int_{\text{I}'_1}dx\,e^{-x^2}\int_{\text{I}'_2}dy\,e^{-y^2}
  \right\}
\end{equation}
with the integration intervals
\begin{eqnarray}
\label{eq:2-23}
 \text{I}_1 &=& [-1/\sqrt{8}\,,\,+1/\sqrt{8}]\nonumber\\
  \text{I}_2 &=&
   \frac{1}{\sqrt{\tau-\tau'}}
    \left[-2\tau'-\tau x-\sqrt{\tau'/8}\,,\,-2\tau'-\tau x+
      \sqrt{\tau'/8}\right]\nonumber\\
     \text{I}'_1&=&
      \left[2\sqrt{\tau}-1/\sqrt{8}\,,\,2\sqrt{\tau}+
       1/\sqrt{8}\right]\nonumber\\
     \text{I}'_2 &=&
       \frac{1}{\sqrt{\tau'-\tau}}
     \left[-\tau x-\sqrt{\tau'/8}\,,\,-\tau x+\sqrt{\tau'/8}\right].
\end{eqnarray}
Now first consider $\tau\approx 1$, $\tau'\gg 1$. Then the first
summand in the jump probability (\ref{eq:2-22}) becomes negligibly
small since $\text{I}_2$ covers a relatively small interval far out
in the wing of the Gaussian integrand; the second summand in
(\ref{eq:2-22}), however, approaches a limit independent of $\tau'$
since $\text{I}'_2\to[-1/\sqrt{8}\,,\,+1/\sqrt{8}]$; therefore, the
jump probability $p(\tau\approx 1,\tau'\gg 1)$ is not small compared
to unity. However, now consider what happens for large times,
$\tau'>\tau\gg 1$; then both summands in (\ref{eq:2-22}) tend to
become vanishingly small, i.~e.~of the order $e^{-4\tau^2}$
as $\tau$ grows large. Indeed, once one has waited several
decoherence times after the preparation of the cat state, jumps
between the macroscopically distinguishable values
$\langle X\rangle_\theta=\pm\alpha$ become exceedingly unlikely. This
behavior is fully borne out by numerical simulations, as exemplified
in Fig.~\ref{fig2} where two ``trajectories''
$\langle X(t)\rangle_\theta$ pertaining to two realizations of the
noise $\theta(t)$ are displayed.

The foregoing reasoning makes for a certain temptation to associate
single realizations of the noise $\theta(t)$ and the accompanying
$\langle X(t)\rangle_\theta$ with single runs of an experiment. We
shall discuss the legitimacy of such an interpretation further below.
It may be appropriate, however, to right away ease the temptation
mentioned by throwing a glance at Fig.~\ref{fig3} which depicts a
simulation of $\langle X_{\pi /2}(t)\rangle_\theta$ for a single
realization of the noise $\theta (t)$. Obviously, in contrast to
$\langle X_0(t)\rangle_\theta$ the mean of the quadrature component
$X_{\pi /2}$ shows no tendency towards choosing one of the two
distinguished values $\pm\alpha\exp (-\frac{\gamma}{2}t)$; even until
times of the order of the mechanical relaxation time $1/\gamma$ the
mean $\langle X_{\pi/2}(t)\rangle_\theta$ stays close to zero before
beginning to display some sizable fluctuations and then eventually
quieting down at zero again for $t\gg 1/\gamma$. This behavior
\cite{CaKoTi,Ca2} of $\langle X_{\pi /2}(t)\rangle_\theta$ is easily
understood by recalling the remark on the dependence of that mean on
the phase $\varphi$ of the local oscillator, made after
(\ref{eq:2-7}); for $\varphi =\pi /2$ one must wait until the overlap
(\ref{eq:2-8}) of the two coherent states $|\pm\alpha\exp
(-\frac{\gamma}{2}t)\rangle$ has grown to values of order unity
before one can expect to see sizable fluctuations of the mean
$\langle X_{\pi /2}(t)\rangle_\theta$ away from zero.

\section{Evolution of cat states according to the nonlinear
Schr\"odinger equation}
\label{sec:evol}
In \cite{GhPeRi,BeSt,GoGr2,GoGr3} the linear stochastic Schr\"odinger
equation (\ref{eq:2-1}) was shown to be equivalent to the following
nonlinear stochastic Schr\"odinger equation
\begin{eqnarray}
\label{eq:3-1}
 d|\phi_\xi(t)\rangle & = & \bigg\{
   -\frac{\gamma}{2} (a^\dagger a-2a\langle X(t)\rangle_\xi+
     \langle X(t)\rangle_\xi^2)dt\nonumber\\
 & & \qquad +\sqrt{\gamma}(a-\langle X(t)\rangle_\xi)d\xi(t)\bigg\}
    |\phi_\xi(t)\rangle
\end{eqnarray}
where the mean displacement
$\langle X(t)\rangle_\xi =\frac{1}{2}\langle a+a^\dagger\rangle_\xi$
is meant with respect to the state $|\phi_\xi\rangle$ itself,
\begin{equation}
\label{eq:3-2}
 \langle X(t)\rangle_\xi = \langle\phi_\xi(t)|\frac{1}{2}(a
 +a^\dagger)|\phi_\xi(t)\rangle.
\end{equation}
The noise increment $d\xi(t)$ represents a Wiener process. The
derivation of the nonlinear evolution equation (\ref{eq:3-1}) from
the linear one, (\ref{eq:2-1}) proceeds in two steps
\cite{GhPeRi,BeSt,GoGr2,GoGr3}. First, one interprets the noise
$d\theta(t)$ in (\ref{eq:2-1}) as output noise driven by a Wiener
input $d\xi(t)$ according to (\ref{eq:2-14}). By then normalizing as
\begin{equation}
\label{eq:3-3}
|\phi_\xi\rangle=|\psi_{\theta(\{\xi\})}\rangle/
 \sqrt{\langle\psi_{\theta(\{\xi\})}|\psi_{\theta(\{\xi\})}\rangle}
\end{equation}
one arrives at (\ref{eq:3-1}). This equation has been employed
previously by several authors (see e.~g.~\cite{WiMi,Ca,GoGr1}).

Due to the equivalence of the linear and the nonlinear stochastic
Schr\"odinger equation and due to the existence of a rigorous
solution of the linear equation, it should also be possible to
construct the exact solution of the nonlinear equation. This has in
fact been achieved recently by Carmichael, Kochan and Tian
\cite{CaKoTi} in the following way. One removes part of the
nonlinearity from the nonlinear
stochastic Schr\"odinger equation (\ref{eq:3-1}) by working with
the non-normalized wave vector $|\psi_{\theta(\{\xi\})}(t)\rangle$
(see (\ref{eq:3-3})), thus obtaining
\begin{equation}
 d|\psi_{\theta(\{\xi\})}(t)\rangle = \bigg\{
   -\frac{\gamma}{2} (a^\dagger a-2a\langle X(t)\rangle_\xi
    dt+\sqrt{\gamma}a d\xi(t)\bigg\}
    |\psi_{\theta(\{\xi\})}(t)\rangle.
\end{equation}
The ansatz
\begin{equation}
\label{eq:3-4}
 |\psi_{\theta(\{\xi\})}(t)\rangle = e^{\alpha^2(e^{-\gamma t}-1)}
 \left(e^{\chi(t)} |\alpha e^{-\frac{\gamma}{2}t}\rangle+e^{-\chi(t)}
 |-\alpha e^{-\frac{\gamma}{2}t}\rangle\right)
\end{equation}
yields an Ito differential equation for the random process $\chi(t)$.
The associated Fokker-Planck equation for the density $P(\chi,t)$
happens to allow for the solution given in \cite{CaKoTi}. After
simplifying as before, i.~e.~$e^{-\gamma t}-1\rightarrow -\gamma t$,
$\langle\alpha|-\alpha\rangle\approx 0$, the solution $P(\chi,t)$
originating from $P(\chi,0)=\delta(\chi)$ takes the form of a sum
of two Gaussians,
\begin{equation}
\label{eq:3-5}
 P(\chi,t)=\frac{1}{2\sqrt{2\pi\alpha^2\gamma t}}\left( e^{(\chi+
 2\alpha^2\gamma t)^2/2\alpha^2\gamma t}+e^{(\chi-2\alpha^2\gamma
 t)^2/2\alpha^2\gamma t}\right)
\end{equation}
which in structure resembles our density (\ref{eq:2-17}) of the
output noise $\theta(t)$. It is most interesting to again see the
width of each of the two Gaussians grow as $\sqrt{t}$ while their
separation grows as $t$, i.~e.~much faster. Moreover, we again find
the separation to become manifest after a time of the order of the
decoherence time $t_{\text{dec}}=1/2\gamma\alpha^2$.

The equivalence of the two versions (\ref{eq:2-1}) and (\ref{eq:3-1})
of the stochastic Schr\"odinger equation is worth one more
illustration. The nonlinear equation (\ref{eq:3-1}) is easily seen
to imply the following equation for the mean $\langle X(t)
\rangle_\xi$ defined in (\ref{eq:3-2}),
\begin{equation}
\label{eq:3-6}
 d\langle X(t)\rangle_\xi = 2\sqrt{\gamma}(\alpha^2-\langle X(t)
 \rangle_\xi^2)d\xi(t).
\end{equation}
The Ito integral of this equation takes the form
\begin{equation}
\label{eq:3-7}
 \langle X(t)\rangle_\xi = \alpha\tanh
  \left( 2(\alpha\sqrt{\gamma})\xi(t)+4\gamma\alpha
    \int_0^t dt'\,\langle X(t')\rangle_\xi\right).
\end{equation}
Obviously, we have arrived at an integral equation rather than an
explicit solution. However, the equivalence of the integral equation
(\ref{eq:3-7}) with the solution (\ref{eq:2-9}) of the linear
Schr\"odinger equation can clearly be seen with the help of
eq.~(\ref{eq:2-14}).

\section{Preparation of a cat state}
\label{sec:prep}
Imagine a spin 1/2 (or equivalently, a two-level atom) prepared
in an eigenstate of the component $S_x$ with eigenvalue $+\hbar/2$.
In the $S_z$ representation that state will take the form 

$\frac{1}{\sqrt{2}} \bigg(|+)+|-)\bigg)$ with
$S_z|\pm)=\pm\frac{\hbar}{2}|\pm)$. Let, on the other hand, an
harmonic oscillator be prepared in the vacuum state $|0\rangle$,
with $a|0\rangle=0$. If we couple the two systems impulsively
according to the Hamiltonian 

\begin{equation}
\label{eq:4-1}
 H(t)=\delta(t)i2(\alpha a^\dagger-\alpha^* a)S_z
\end{equation}
we will produce the composite state
\begin{equation}
\label{eq:4-2}
  \frac{1}{\sqrt{2}}\bigg(|\alpha\rangle|+)+|-\alpha\rangle|-)\bigg)
\end{equation}
in which the coherent state $|\alpha\rangle$ of the oscillator is
correlated with the ``up'' state $|+)$ of the spin while the coherent
state of the opposite amplitude, $|-\alpha\rangle$, is correlated
with the spin-down-state $|-)$. Needless to say, the composite state
(\ref{eq:4-2}) is not a cat state of the oscillator, even if the
amplitude $\alpha$ is large.

In order to proceed towards preparing a cat state we let the spin
be exposed to a magnetic field in the $y$ direction such that spin
states are transformed as
\begin{equation}
  \label{eq:4-4}
   |\pm)\to e^{-i\pi S_y/2\hbar}|\pm)=
    \frac{1}{\sqrt{2}}\bigg(|+)\pm|-)\bigg).
\end{equation}
This corresponds to a spin rotation by $\pi/2$ about the $y$ axis.
The composite state (\ref{eq:4-2}) is thus transformed into
\begin{equation}
 \label{eq:4-5}
  \frac{1}{\sqrt{2}}\Bigg\{ 
    |+)\frac{1}{\sqrt{2}}
    \bigg(|\alpha\rangle+|-\alpha\rangle\bigg)
     +|-)\frac{1}{\sqrt{2}}
      \bigg(|\alpha\rangle-|-\alpha\rangle\bigg)
       \Bigg\}\,.
\end{equation}
Now both eigenstates of $S_z$ are correlated with cat states of the
oscillator, if $|\alpha|\gg 1$. We may finally imagine a measurement
of $S_z$. Every time we find $S_z=+\hbar/2$ we know that the
oscillator is prepared in the cat state 

$\frac{1}{2}(|\alpha\rangle+|-\alpha\rangle)$.

In principle, the thought experiment just sketched can be realized
by playing with two level atoms traversing microwave or
optical cavities. Care would have to be taken that the two-step
preparation takes less time than a decoherence time
$t_{\text{dec}}=1/2\gamma|\alpha|^2$ where $\gamma$ is the damping
constant of the cavity within which the cat state is to be
produced.

\section{Conclusion and discussion}
Mostly employing a linear version of the stochastic Schr\"odinger
equation we have presented a rigorous description of the fate of an
initial state of the cat type for the damped harmonic oscillator.
While our description is stochastically equivalent to one using the
master equation for the density operator, it has the additional
appeal of being more closely related to individual runs of an
experiment rather than to ensembles of such. Indeed, the fate of the
cat state described here makes for a certain temptation to associate
a single realization of the noise $\theta(t)$ and the accompanying
mean displacement $\langle X(t)\rangle_\theta$ with a single run of
an experiment. In fact, such an interpretation is known to be
legitimate only with respect to balanced homodyne experiments where
the output of the quantum oscillator is put to interference with a
local oscillator of large amplitude and stable phase $\varphi=0$;
sharp values are then in effect recorded for the output noise
$\theta(t)$ of the quantum oscillator.

Inasmuch as the mean displacement $\langle X(t)\rangle_\theta$ tends
to assume, for times in the interval $1/\gamma\alpha^2\ll t \ll
1/\gamma$, either one of the two values $\pm\alpha$, one may even
feel inclined towards regarding the displacement $X$ itself as a
measured observable. The exceedingly small likelihood of jumps
between the two preferred values $\pm\alpha$ during the time interval
mentioned does permit some such indulgence. One must keep in mind,
though, that both for early times, $t\approx 1/\gamma\alpha^2$, and
large times, $t\approx 1/\gamma$, the displacement cannot be
considered as sharp in an individual run of the experiment; nor can
the quadrature component $X_{\pi /2}(t) $ at any time.

Some further caution is indicated against hurried conclusions for
quantum measurement theory. The heat bath providing the damping of
the quantum oscillator is, strictly speaking, not the one needed for
``objectivation'' of sharp ``pointer readings''; nor is the
displacement $X$ really playing the role of a pointer variable. In
our context the quantum oscillator rather acts as a measured object
and objectivation only arises in the photodetector which generates a
macroscopic electric current. The pointer variable with respect to
which the detector must secure decoherence, i.~e.~objectivation, is
just the output noise $\theta(t)$ for which sharp values are
recorded. A striking manifestation of the
potential ``nonobjectivity'' of the interaction of the oscillator
with the heat bath may also be seen in the sensitive dependence of
the statistics of $X_\varphi(t)$ on the phase $\varphi$ of the local
oscillator, discussed in Section \ref{sec:linear} and already noted
in \cite{CaKoTi,Ca2}. Astounding as that sensitivity may be, it does
not make for worries in the context of quantum measurement theory;
while the phase of the local oscillator does not influence the
interaction of the quantum oscillator with its damping reservoir,
that phase does enter the oscillator observable coupled to the
detector, i.~e.~the ouput noise $\theta(t)$. The interested reader
is referred to Zurek's discussion of ``pointer bases'' with respect
to which fast decoherence takes place \cite{Zu}.

\acknowledgements
We gratefully acknowledge support by the Sonderforschungsbereich
``Unordnung und grosse Fluktuationen'' of the Deutsche
Forschungsgemeinschaft. We thank Howard Car\-michael for sending us
preprints of his work prior to publication.

\newpage
\begin{figure}
\caption{Marginal density of the output noise $\theta$ for different
times $\tau$ measured in units of the decoherence time for parameter
values $\gamma=0.1$, $|\alpha|=3$.}
\label{fig1}
\end{figure}

\begin{figure}
\caption{The mean $\langle X_0 (t)\rangle_\theta$ for two runs of the
nonlinear stochastic Schr\"odinger equation with initial amplitudes
$\alpha=\pm 3$. The dashed lines show the mechanical relaxation
$\pm\alpha\exp(-\gamma t/2)$ of the amplitudes of the coherent
states. In (a), the approach to $+\alpha\exp(-\gamma t/2)$ at
$t\approx 0.75~t_{\text{dec}}$ is not quite close enough to produce
definite locking which occurs only at $t\approx 2.6~t_{\text{dec}}$.}
\label{fig2}
\end{figure}

\begin{figure}
\caption{The mean $\langle X_{\pi /2}(t)\rangle_\theta$. Again,
$\alpha = \pm 3$.}
\label{fig3}
\end{figure}

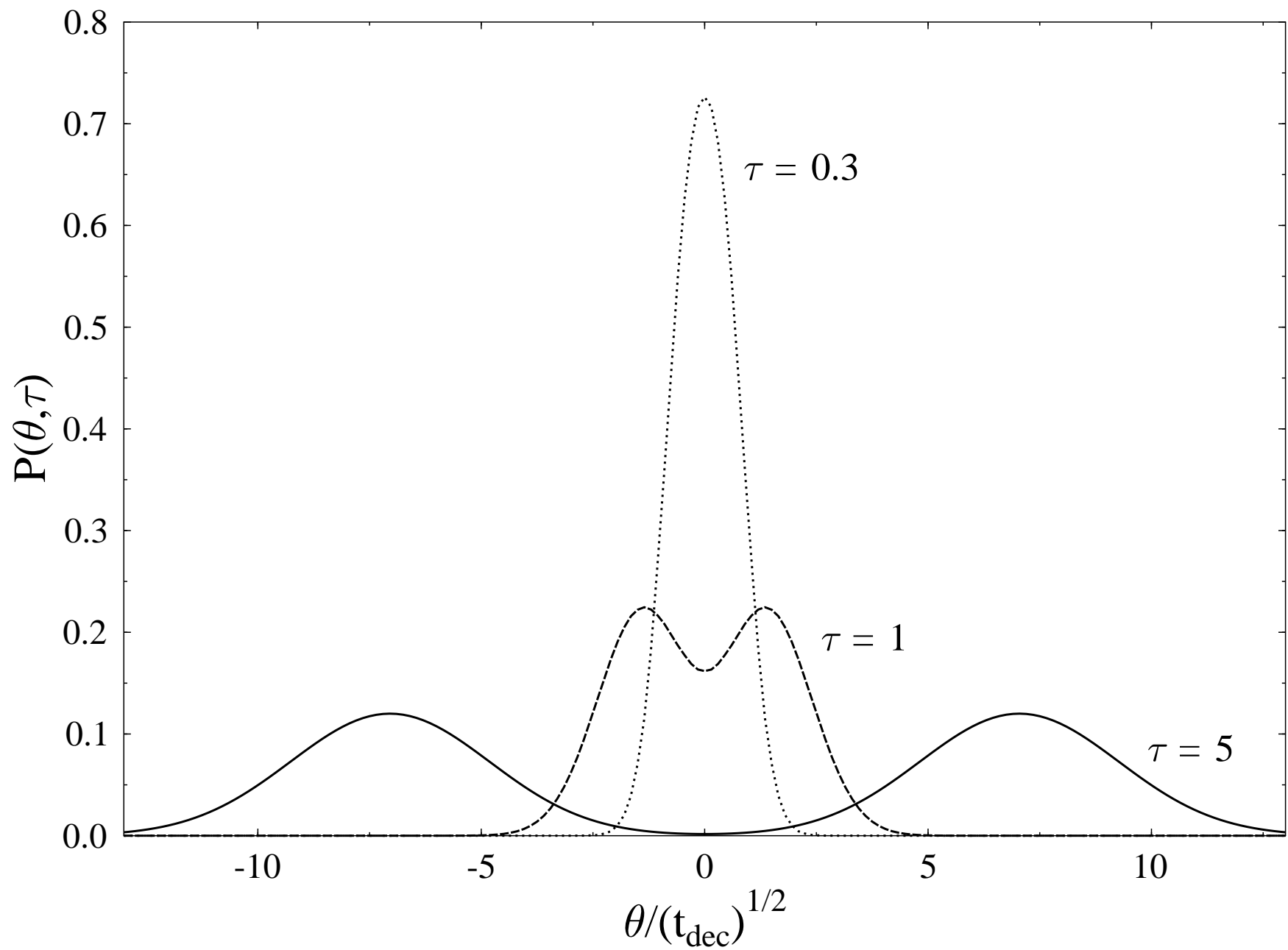
\end{document}